\documentstyle[prb,aps,twocolumn,amssymb,amsfonts,epsfig]{revtex}

\begin{document}
\draft \flushbottom

\wideabs{

\title{
Electronic structure and magnetism in slightly doped SrB$_6$.
}

\author{T. Jarlborg}
\address{
DPMC, University of Geneva, 
24 Quai Ernest-Ansermet, CH-1211 Gen\`eve 4, Switzerland
}
\date{\today}
\maketitle

\begin{abstract}
Spin-polarized band calculations for supercells of SrB$_{6}$, where a La-, In- or Al- impurity
or a vacancy is replacing one Sr, are performed within
the local spin density approximation. Moderately large cells with 8 formula units (56 atoms) 
are studied for all dopings
and large ones with 27 formula units (189 atoms) for the case of La-doping.  
The undoped system has a vanishing density-of-states (DOS)
at the Fermi energy (E$_F$), while the additional La-d band makes E$_F$ to enter the bands above the gap.
An Al (or In) impurity has the opposite effect, with a rigid-band like shift of E$_F$ to 
below the gap. In the former case,  the addition of a d-electron
makes a local, impurity-like 
modification of the electronic structure close to the La atom. As has been shown in a previous
publication {[}Phys. Rev. Lett. {\bf 85}, 186, (2000){]}, it can lead to a weak ferromagnetic state. 
This result is shown to persist
at even lower doping, by the 189-atom cell calculations, with a moment of the order 0.1 $\mu_B$
per La impurity. An addition of a p-electron, by an In-impurity,
makes no similar effect.
The DOS at E$_F$ and the Stoner factors are in all cases low, and would not suggest ferromagnetism alone. 
The  magnetic state can be understood from a charge transfer and an additional gain in
 potential energy,  as a spin-splitting is imposed. A DOS with a localized impurity band 
is essential for this effect. Different models with modified localization of the La-band are made
in order to show the correlation between charge transfer and the size of the magnetic moment.


\end{abstract}


}

\section{Introduction.}
The structure of hexaborides, XB$_6$, is simple cubic, with one large atom (X = Eu, Ca, Sr or Ba) in the corner, 
and a B$_6$ octahedron
in the center. It can be viewed as the CsCl-structure, where the B$_6$-octahedron would be a pseudo-atom
in the center
and the X-atom in the corner is the second atom.
 The internal parameter, $z$, is a measure of the size of the octahedron. It
is the shortest distance from one of the B-atoms
at the edge of an octahedron to the cube surface. Its value is about 0.2 of the lattice constant, but
it varies from system to system.
The hexaborides have been studied mostly because their intriguing 
properties showing signatures of both metallic and semi-conducting behavior \cite{aron,ott1,deg}.
Recent experimental works on La-doped hexaborides like La$_x$Ca$_{(1-x)}$B$_6$
show unexpectedly  a very weak ferromagnetic state 
that persists up to large
temperatures, and for very low $x$, about 0.005  \cite{you}. Independent measurements confirm
these findings in part, but they also show that the results can span from weak ferromagnetism 
to strong paramagnetism
and that Al impurities may conceal some of the intrinsic properties \cite{tera}.

 Ferromagnetism in the La-doped hexaborides XB$_6$, X=Ca,Sr or Ba, is surprising because the
density-of-states (DOS) at the Fermi energy, $N(E_F)$, and the Stoner
factor,  $\bar{S}$, are expected to be small in this family of compounds. 
Band calculations \cite{mass} in the  local density approximation (LDA) \cite{lda}
on SrB$_6$ and CaB$_6$ confirm that the DOS is low, with $E_F$ falling in a
valley of the DOS, and spin-polarized calculations for EuB$_6$ show a very
weak spill-over of the spin density from the localized Eu-f spin to the rest
of the valence band. The exact behavior of the band overlap at the X-point has been monitored
in several band calculations \cite{hase,mass,rod,tro}. It has been found that this overlap, which
determines if the undoped system is metallic or semi-conducting, depends on the structural parameter
$z$. In fact, the overlap is more sensitive to $z$-variations than to variations of 
the lattice constant \cite{mass}. A recent band calculation \cite{rod} determined the fine details of 
the bands near
$E_F$ and the Fermi surface (FS), and concluded that they are in good agreement with 
de-Haas-van-Alphen
measurements \cite{dhva}. An experimental  determination of the band dispersion the near the X-point,
by photoemisson  \cite{arpes}, is in fair
agreement with the theoretical predictions for the existence of a X-pocket in the bulk, but a different
interpretation is made for bands in the surface region. 
 
Electronic structure calculations based on the local (spin) density
approximation, either by spin-polarized calculations or by
calculations of the Stoner factor $\bar{S}$, are in general able to detect if a
system is ferromagnetic or not \cite{hunt,jan,c15}. Because the Stoner factor for La$_x$Sr$_{(1-x)}$B$_6$
is  far below the limit for magnetism \cite{hexlet},  it is tempting to suggest that the
observed magnetism is not a band magnetism, but something else. 
Electron-hole interaction for an asymmetric position
of E$_F$ relative to the gap has been proposed to lead to excitonic magnetism \cite{ric}. Spontaneous
polarization of a dilute electron-gas has also been suggested \cite{you,ort}.
However, for Ce
it is found that the Stoner factor underestimates the tendency towards magnetism,
because the stabilization energy for a magnetic state comes also from the Coulomb energy 
in addition to the exchange energy \cite{tj98}. The effects can be larger in a compound 
as long as the different sub-bands and
the ionicity of the system show favorable conditions.  
In a recent work we showed that LaSr$_7$B$_{48}$ is likely to be such
a system \cite{hexlet}. Here we extend these calculations to larger unit cells, to different types
of dopants, and give a more
complete discussion of the results.

\section{Theory.}

\subsection{Band method.}

Electron energies are calculated 
by the linear muffin-tin orbital (LMTO) method \cite{lmto} using the LDA.
 The internal parameter $z$, is in these calculations 0.215 of the lattice constant. 
 The lattice constant is fixed at 7.93 a.u. in each calculation.
 Additional 'empty' spheres
are inserted at the cube edges, between the Sr-atoms, in order to account for non-sphericity
of the potential in the open part of the structure. 
Calculated results are presented for two unit cells: One 
 supercell of 8 elementary cubes (2 cube lengths in x,y and z-direction), containing totally 80 sites, 
56 atoms and 24 empty spheres. A second set of calculations was made using a larger super cell
containing 27 elementary cells (3 cube lengths in each direction), with totally 270 sites (189 atoms
plus 81 empty spheres). These two cells allow to study doping concentrations $x$ of 0.125 and 0.037, respectively.
 Nonequivalent sites are treated independently.
One basis set for the 80-site cell
unit cell includes s,p,d for the atoms, and s,p for empty spheres. The results from these
calculations can be found in ref. \cite{hexlet}. A reduced
basis set for the large 270-site calculations includes s and p states for the B-atoms 
and  only s-states for the empty sites in the direct basis.  
We include the B-d and empty-p "tail" states in the reduced basis
calculations. This means that the d-states for B atoms, for example, are constructed from 
tail states coming from neighboring sites. This gives a good description of
tail states like B-d and p-states on the empty sites, close to what was obtained from the large basis,
 without having to increase the size of the
eigenvalue problem. Intrinsic "on-site" states, like the high energy La-f state or  B-d
states, can not be
constructed from tail states in this way, but these states are unoccupied and not important in
this study. 
Test calculations for the smaller of the
cells using this smaller basis showed very similar results as the ones from using the
large basis, when the linearization energies (E$_{\ell}$) of the La-band are increased upwards about 0.2 Ry.

In the 270-site calculations, where we employed the same basis (with tail states)
and the same E$_{\ell}$'s, which in the small basis 80-site calculation reproduced well the results
of the older large basis calculation. In this way we believe that the results for the large 270-site
cell are realistic despite the use of a minimal basis set.
 No lattice relaxation around the impurities
is considered. The number of k-points varies from
10 to 20 irreducible k-points for the unit cell of 80 sites, and from 4 to 10 for the 270 site cell.

\subsection{Stoner models.}

The original Stoner model \cite{hunt} can be applied together with ab-initio paramagnetic 
 band results in order to study the spin susceptibility or
the onset of magnetism  \cite{jan,c15}. 
 An applied exchange splitting, $\xi$, leads to a transfer from minority spins
to majority spins, and a loss 
in kinetic energy $\Delta K = N \xi^2$, where $N$, the DOS
at $E_F$, is assumed to be constant. The same spin transfer leads to a gain in
exchange energy $\Delta E_x = N^2 I_s \xi^2$, where  $I_s$ is
 the exchange integral \cite{jan}.
If a partial DOS is not constant near $E_F$, there will be a possibility of charge transfers, $\Delta q_{\xi}$
and $\Delta Q$, as function
of $\xi$ and temperature, which make additional contributions to the total energy. It is possible to
identify one sub-band, the La-impurity (mainly d) band, which has a large derivative of its DOS, $N'_{La}(E_F)$.
The total DOS at $E_F$ is $N=N_{La} + N_v$, where $N_v$ is the rest of the DOS. 
 By neglecting all other derivatives one can make a
model of the charge transfers as function of $\xi$ and $k_BT$. By taking into account the energy associated
with the charge transfer, $U_0$,  one obtains a modified
Stoner criterion. This is made in detail previously \cite{tj98} and the new Stoner criterion is:

\begin{equation}
\label{eq:msto} 
\bar{S}_U = N I_s + U_0 N'_{La} N_v /N^2  \geq 1    
\end{equation}

This modified Stoner criterion is useful for an understanding of why a system can be
pushed towards magnetism by Coulomb energies. But, as will be discussed in section 3.2, 
the effects are too strong to
use eq. \ref{eq:msto} for a quantitative prediction of a magnetic state.

\section{Results.}

\subsection{Bandstructures.}

The bandstructure for undoped SrB$_6$ agree well with the one in ref. \cite{mass}.
The Fermi energy falls in a low DOS region between the B-p and Sr-d band regions, where a small gap can
appear. The B octahedron behaves
partly as one identity, as a large pseudo atom. This is seen in a narrow band about 0.9 Ry below E$_F$.
There is only one such band per basic unit cell, although in reality it is of B s- and p-character. This band is
separate from the other B s- and p-bands. Its presence could be viewed as one single s-band from
each B$_6$ pseudo atom within a CsCl-structure, where Sr is one atom and B$_6$ the other one. However, this is far
below E$_F$ and this way of looking at the B-octahedron is probably not useful for the bands near E$_F$ or
 for the physical properties
of doped hexaborides.

\begin{figure}[tb!]
\leavevmode\begin{center}\epsfxsize8.6cm\epsfbox{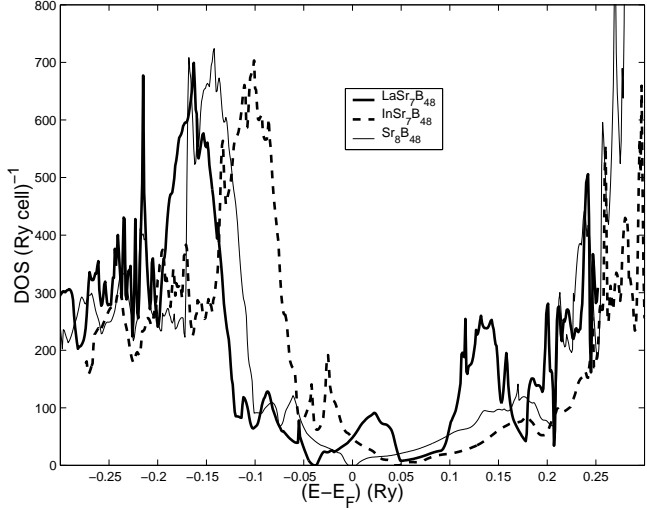}\end{center}
\caption{\label{fig:dopdos}
Paramagnetic, total DOS for LaSr$_7$B$_{48}$ (bold line), InSr$_7$B$_{48}$ (broken line) 
and undoped Sr$_8$B$_{48}$ as obtained from tetrahedron integration for
20 k-points using the small basis. The zero energy is at $E_F$ for each case. Note that In-doping
leads to a {\it downward} rigid-band like shift of E$_F$ on a DOS which resembles much that of the
undoped material. 
}
\end{figure}

The effect of doping, where one Sr is exchanged with another atom, show some unexpected result depending on
the dopant. Fig \ref{fig:dopdos} shows the DOS for undoped, for La-doped and In-doped SrB$_6$, all for
$x$=0.125 (small 80-site cell, small basis). La has a d-band coming in just above E$_F$. 
In this energy range there are very few B and Sr bands for the La-d band to hybridize with. 
The La-d band can hybridize between La neighbors if the La
concentration is large, but if there are only few
La atoms, the La-d band will localize and becomes  narrow.
Indium has one additional valence electron compared to Sr, exactly
as La has. But the additional electron in In is a p-electron, and in contrast to the d-band for La,
the In-p band hybridizes well with the B-bands below E$_F$. The new In-band below E$_F$ can accommodate
two electrons, but In has only one additional electron compared to Sr. 
The result is that E$_F$ will move downwards compared to the rest of the bands. Thus,
as seen in fig. \ref{fig:dopdos}, the effects of In and La doping are opposite. Furthermore, with La there
is a large increase of the local La-DOS, while the DOS on the In-site and a Sr-site are very similar.
 The resulting DOS from In doping is very similar to the one from Al (or even B) doping,
at least as long as lattice relaxation can be ignored. It has been reported that Al can penetrate the
hexaboride material when the aluminum flux-growth technique is used to prepare single crystals \cite{tera}.
The calculated results for single In or Al impurities shown here, indicate that the effects of Al
contamination should be very different from those of La impurities. The effects could even appear quite
undetected compared to undoped hexaboride, since the main result is a rigid-band like downward shift
of E$_F$ on a uniform DOS. Theories of excitonic magnetism are based on rigid-band assumptions of
the undoped band structure \cite{ric}. Such theories should therefore apply to Al (or In) doping, but not easily 
for La-doping, since the latter is not rigid-band like. However, according to our results, the effect of
the p-electron doping in this material is oddly transformed into hole doping.  The charge distribution is
such that about one half of the additional electron from In stays on the site, a bit more than one half is
added close to the impurity. At the most distant sites there is slightly less electronic charge
compared to the values for the undoped case, which presumably is an effect of the shift of E$_F$. 

The paramagnetic (non-spinpolarized) DOS from the calculation for LaSr$_7$B$_{48}$ is shown in
fig. \ref{fig:dopdos}. The La band forms a separate band near $E_F$ below the rest of the Sr-d band,
making the real space distribution to be localized near the La impurity from this band.
The total LaSr$_7$B$_{48}$ DOS at $E_F, N(E_F)$, about 48 states/cell/Ry, is concentrated on
the single La site, with $N_{La}(E_F) \sim$ 19 states/cell/Ry, cf Table 1. On the remaining 55 sites, each Sr has
on average 1.0 and each B roughly 0.4 states/cell/Ry, respectively.
By comparison, a rigid-band model obtained by adding one electron to the band structure without the La
impurity  gives very different values: Of the total DOS, 32 states/cell/Ry,
each Sr has 1.4 and each B 0.25 states/Ry.   
Clearly, La behaves like a localized impurity site with a
very large local $N_{La}(E_F)$, much larger than on Sr. 
 The DOS values are included in Table 1.
 
The additional valence electron remains
located on the La. 
By comparing the charges in the undoped and doped supercell one finds that
the remaining Sr$_7B_{48}$-spheres receive no additional charge, cf. Table 2.  The difference in charge between 
the two systems is never larger than 0.01 electron per Sr or B site. The derivative of the DOS becomes very large, 
about 1400 states/cell/Ry$^2$,
from which 700 comes from the La site, compared to about 30 for each Sr site when the rigid-band model is
applied to the undoped bandstructure. This asymmetric and uneven distribution of the DOS near $E_F$ 
is a condition for real space charge transfers as T increases and as an exchange splitting
is imposed. The transfer goes in both cases to the La site from the rest of the system.

 Next, for the calculations with the large 270-site cell only the case with a La-impurity was considered, 
 LaSr$_{26}$B$_{162}$. 
 The total and partial La-DOS near
 E$_F$ are shown in fig \ref{fig:dos270}, together with the corresponding DOS from the 80-site calculation.
 Both results are calculated with the same basis. 
 The DOS functions from the two calculations are divided by 27 and 8, respectively. This makes
the units of the DOS states
 per formula unit (f.u. of SrB$_6$), and the two cases can be compared on the same diagram. The number of k-points,
 20 for the small and 10 for the large cell, reflects a larger density of points in the reciprocal space
 for the 270-site calculation. It
 is seen that the La-band is more separated from other bands and more localized in the large cell. 
 The value of the total N at E$_F$  (6 states per Ry and f.u.)
is quite similar in the two cases, but the local DOS on the La-site is much larger in the largest cell. The 270-site
cell has a large number of sites far from the impurity with low DOS, almost as for the undoped material. The
number of such sites is comparatively larger than for the 80-site cell.
 But the very large DOS at the impurity site compensates for the low DOS elsewhere, and 
the average DOS per f.u. becomes similar for the two cells. The local La-DOS, which in the small cell is about 19, 
is 55 in the large cell, cf Table 3.

 A comparison of the DOS near E$_F$ 
with the ones from the small cell, reveals more rapid variations of the DOS structures. 
The La-band is narrower, with
the total DOS going from zero to very large values within a much narrower energy interval than
in the case of the 80-site cell. The effect is larger DOS derivatives, mainly steming from La, but also
larger sensitivity of the charge transfers and their T-dependencies.
The amount of charge transfer as function
of $k_BT$ (and $\xi$) is typically 3-5 times larger within the large
cell, but the non-linearities are large and a small shift of E$_F$ can make a large difference. It is
therefore not meaningful to quote values of $N'$ in this case.
At the most distant (from the La) B sites,
the local B-DOS is small, but still about 50 percent larger than what is the case for the undoped DOS, 
when a rigid band shift has been done
to account for one additional electron. For the distant Sr-sites, the local DOS tend to be close to the value
of the undoped DOS with the proper rigid-band shift of E$_F$, 
about 0.7 per Ry. The first 1-2 layers of Sr and B surrounding the La impurity, have approximately
doubled their local DOS compared to the undoped values. 

 Despite the large number of
atoms separating the La in the largest cell (188 atoms, 3 types of Sr and 9 types of B), and despite the fact that
the additional La electron stays localized near the impurity, there is still an influence on the electronic
structure far
from the impurity.  Their DOS values are 
larger than the ones of the undoped DOS, which are obtained after a  
 rigid-band shift of E$_F$ to account for an additional electron. An
even lower impurity concentration seems to be required for finding the electronic structure far from
the impurity, almost
"insulating" like for the undoped case. In such a case, the interaction between local moments
would effectively be cut-off.


\subsection{Magnetism.}

We first present the results using the 80-site cells, mostly using the small basis set as explained above.
Spin-polarized calculations for undoped SrB$_6$
do not give stable moments of significant size. An applied field is used to get a moment of about 0.5 $\mu_B$ per
cell, but this moment decreases steadily during the iterations when the field is removed. The same situation holds
for InSr$_7$B$_{48}$. A number of calculations were also attempted where one Sr was replaced with an 'empty' site.
This was motivated by the reports that a kind of "self-doping", by Sr vacancies might be present because
some seemingly undoped samples show the weak magnetism \cite{you}. The unpolarized calculations for the cases with
a Sr-vacancy showed the effect of charge transfer as function of T, so that as T increased there was a larger charge
on the remaining Sr sites. However, the moments always decreased (to below 0.005 $\mu_B$) 
in the spin-polarized calculations,
also for the case where a slight B-relaxation towards the missing Sr-site was attempted. Thus, in the following 
we discuss only
La-doping, since undoped, In- (or Al-) doping, and Sr-vacancy "self-doping" all failed to show magnetism.

The 80-site cell calculation of LaSr$_7$B$_{48}$ using 20 k-points
and the large basis,
finds a small moment of 0.10 $\mu_B$/cell, which is 
 about 10 mRy/supercell lower in total energy than the paramagnetic solution \cite{hexlet}. 
If fewer k-points are used,
one finds that the (paramagnetic) DOS and its energy derivative near $E_F$ are larger, cf Table 1. One case
with the small basis is also shown. The moments are not quite as large as with the large basis, but
qualitatively the results are the same.

A number of different calculations with the 80-site cell and the small basis are done where the linearization
energies are varied in order to obtain different values for N(E$_F$) and its derivative. All these different cases
are not presented in detail, but the trend is that when the local La-DOS and its derivative at E$_F$ are large,
then the moments will be large (larger than about 0.1 $\mu_B$ per cell) in the spin-polarized calculations. 
Thermal broadening will usually oppose a large moment. But there are a few exceptions, when the
energy derivative of the DOS is varying around E$_F$, showing that the sharpness of the DOS is not
always close to E$_F$. Temperature dependent charge transfers also show deviations from a simple T$^2$
behavior.
The moment is always largest on La, with more than 50 percent of the
total moment on and near the La impurity. The moment on B-sites
far from the impurity is negligible. This fact and the uneven distribution of local DOS show that
a rigid-band picture of dilute La-doped hexaborides does not apply. The bands near $E_F$
are modified by the spatially non-uniform perturbation of the potential around the impurity site.

 The calculated Stoner product $N I_s$ is 0.21 \cite{hexlet}, $N'_{La} \sim$ 730 per Ry$^2 \cdot$ cell, 
 N $\sim 48$ and $N_v \sim 29$ (per Ry $\cdot$ cell), 
to give
$\Delta q_{\xi} \approx 440 \xi^2$  and $\Delta Q = 720 (k_BT)^2$,  with $\xi$ and $k_BT$ in Ry.
With $U_0$ calculated from T-dependent variations of the total energies to be about 1.5 Ry/el./La, 
the Coulomb term in eq. \ref{eq:msto} becomes very large, about 14. 
However, even if the Coulomb
energy represents an important correction in this case, there are several
facts indicating that other energies are involved, and that the response of the system (during a self-consistent,
spin-polarized calculation) will prevent the system to absorb all of the energy. 
First, an exchange splitting $\xi$ which is not uniform over all sites
will modify the energy. The spin-polarized calculation 
shows that $\xi$ is finite (3-5 mRy) only close to La. Second, although the amount of charge transfer
due to $\xi$ and $k_BT$  is similar, there are long tails of 
the Fermi-Dirac distribution which make the radial dependencies of $\Delta q_{\xi}$ and $\Delta Q(T)$ different
and make $U_0$ non-transferable from one case to the other.
Third,
 the effective derivative of the La-DOS, $N'_{La}(\epsilon,T)$, (taken as $\int (-\frac{df}{d\epsilon})
 N'_f(\epsilon) d\epsilon$), is T-dependent, because the bare DOS does not have the same 
 large derivative over a wide energy range. 
 The mechanism of charge transfer and magnetism is therefore expected to 
 depend on T. 
This is confirmed in the spinpolarized results. Usually, by raising $k_BT$, the magnetic moment goes down.
But a few cases are found where the moment is largest for $k_BT$ about 3 mRy. The reason is that
the sharpest DOS structure is a bit beside the position of E$_F$ in that case.

Calculations of total energies converge slowly and a comparison between paramagnetic
and spin-polarized results are not 
made in all cases. From the converged cases it is seen that the total energies for magnetic solutions are lower than 
the non-magnetic ones.  This
difference is largest when the moments are largest, 10-14 mRy/cell for a total moment of 0.10-0.19 $\mu_B$. 
For the cases with moments smaller than 0.05 $\mu_B$, the difference is only 0-2 mRy.

\begin{figure}[tb!] 
\leavevmode\begin{center}\epsfxsize8.6cm\epsfbox{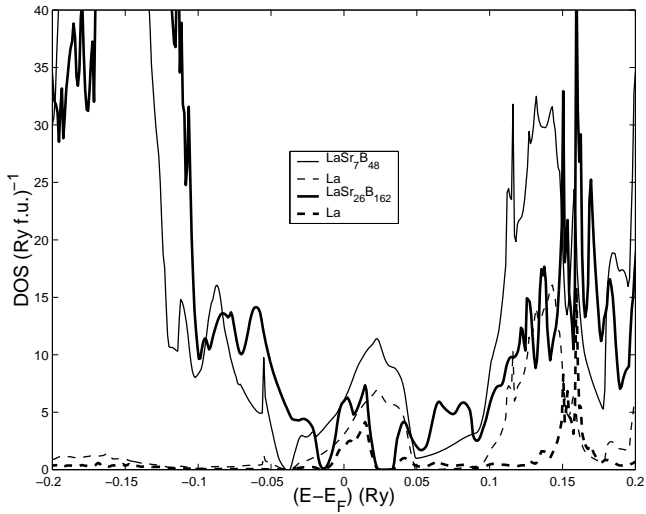}\end{center}
\caption{\label{fig:dos270}
Paramagnetic, total and partial La-DOS (including the DOS from the 6 closest empty spheres)
near $E_F$ for LaSr$_7$B$_{48}$
and LaSr$_{26}$B$_{162}$ as obtained from tetrahedron integration for
20 and 10 k-points, respectively. The zero energy is at $E_F$ in both cases. Note that the DOS is per formula
unit (SrB$_6$), so that the total DOS is divided by 8 in the former, and by 27 in the latter case, respectively.
Thus, the partial La DOS is much larger in the large cell (about 55 per La) compared to in the small cell 
(about 20 per La).
}
\end{figure}

 Two sets of spin-polarized calculations were done for the 270-site cells, both with 4 and 10 k-points. One
used the same linearization energies, which for the 80-site cell reproduced well the result with
the large basis. It is the paramagnetic DOS functions from these parametrizations which are shown in  
fig. \ref{fig:dos270}. The DOS values at E$_F$ and magnetic results are displayed in Table 3. 
The other set use different linearization energies, which
for the paramagnetic 80- and 270-site calculations lead to increased DOS and derivative of the DOS.
The spin-polarized results using 10 k-points show a much larger moment for the 
latter case compared to the first set. The
reason is that there is a region of a rather flat (or less steep) DOS  close to E$_F$ in the first case,
as can be seen in fig. \ref{fig:dos270}. The bold line of the DOS from the 270-site cell is seen to follow
the thin line of the 80-site cell, but only within a few mRy of E$_F$. Further below E$_F$ the curve is
again steep, as is the case for the other set of calculations for the 270-site cell. The result is that
the charge transfer is lower in the former set of calculations. However, this difference between the two sets
of calculations diminishes at larger T when the steep region of the DOS a bit aside of E$_F$ becomes involved.
This is consistent with the T-dependence of the moment, since
the largest moment in the first set of calculations shown in Table 3, is when $k_BT$ is 4 mRy. At low
T ($k_BT \sim$ 0.5 mRy), the moment
decreases further. The T-dependence of the moment is in general opposite for the other cases, 
meaning that the moments
become very small when $k_BT \sim$ 3 mRy.  For the other set and in most cases with the 
80-site calculations the moments
are largest for the smallest $k_BT \sim$ 0.5 mRy.

 When the virtual crystal approximation is applied to this case, to remove 0.5 electrons from the system 
(and 0.5 nuclear charges from the Sr sites), one finds that E$_F$ falls closer to the steep left-hand side
of the La-DOS, cf. fig  \ref{fig:dos270}. The objective with this model calculation is to see if the 
new conditions make the moment
to increase, as was expected from the previous discussion. It should also approach the situation with 
two-component doping, where Al-doping (and a rigid-band shift of E$_F$) in addition to La-doping could 
influence the magnetic state. The result
is that the spin-polarized results converge towards
a larger moment, about 0.09 $\mu_B$, but now only for low T. At large $k_BT \sim 3$ mRy,
the moment shows a steady decrease.
These T-dependent results corroborate the finding that there is
a close connection between sharp DOS structures, charge transfer and magnetic moments.

 The results in Table 1 and 3 show that the moments per impurity are comparable for the small and the large cell, 
about 0.1 $\mu_B$ per
cell. The correlation between large derivative of the DOS and the moment, is seen in the results, both
from the variation of the number of k-points and from the different choice of linearization. The largest moment,
0.25 $\mu_B$/La is only slightly larger than the largest moment in the 80-site calculation, 0.19  $\mu_B$/La.
This is despite the much larger derivative of the DOS from the results with 270 sites. As was discussed above,
the reason could be that the large cell has a comparatively larger undoped bulk, which opposes polarization.

\subsection{Discussion.}

The results in ref. \cite{hexlet} for an 80-site cell, showed a stable ferromagnetic state, and
 the moment remains centered around the 
La impurity. Together with the paramagnetic results, one can conclude
that the magnetism is correlated with the derivative of the DOS on La,
and that the magnetic state is partly stabilized by a charge localization on the La impurity. This correlation
becomes more evident in the continued calculations using smaller basis, which are modified in order to
create different conditions for large or small charge transfer effects. The results from the 270-site calculations
show that the conditions for charge transfer to the La impurity are reinforced, because the La-d band becomes
even more narrow and localized due to the increased La-La distance in the large cell. The magnetic moments
per La-impurity are comparable in the two cells though, probably because of the larger "bulk"
between the impurities, which is undoped and counteracts magnetism better in the large cell.
On the other hand there is no magnetic state for the In or Al impurity, for the Sr-vacancy or for
the undoped SrB$_6$. This is despite a certain pre-condition for a charge transfer from B towards remaining
Sr in all these cases, but apparently it is too weak to produce an itinerant magnetic state, at
least in the 80-site cell. The weak transfer towards
Sr is present even in the La-doped cases, and therefore it reduces the (larger) transfer towards 
La. This competition could reduce the amplitude of the La-moment and even prevent magnetism
if it became larger. One could speculate that Al substitutions on Sr sites play a role of a charge reservoir 
in this situation. Namely,
as was noted earlier, an Al atom on a Sr-site makes no large effects on the DOS except for a rigid-band like
downward shift of E$_F$. No explicit calculations of having both La- and Al-substitutions
were done, but taken separately these results show impurity band and rigid-band behavior, respectively. 
A varying Al content in a La-doped material should therefore modify the position of
E$_F$ relative to the narrow La-d band, and  create more or less optimal conditions
for the charge transfer around the La-impurity. By tuning with Al-doping to find the largest DOS derivative on
the La-band, one should be able to optimize the magnetic moment.

The non-rigid band like electronic
structure around the dilute La impurity shows up in the distribution of the ferromagnetic moment.
According to these results, it is striking that La and the nearest sites takes the major part
of the total moment. B-sites remain essentially non-polarized. The La-site acts as an impurity
and will be a perturbation for excitons. Ferromagnetism has been proposed to be caused by 
the excitonic mechanism \cite{ric}, since the undoped band structure show some favorable conditions for this.
However, it is seen that with La-doping there are large modifications of the undoped bandstructure, and a
rigid-band assumption is not at all appropriate. Rigid-band like conditions occur for In-, Al-, to some extent for
"vacancy"-doping, with E$_F$ lower than for the undoped case. 
 Theories of spontaneous ferromagnetism of the dilute electron gas
have also been proposed \cite{ort,you}. However, an unrealistically large electron-gas parameter $r_s$ is
necessary for this ($\geq$ 30), whereas the calculated  $r_s$ is smaller than about 1.5.

Some experiments can be suggested to verify if the La-band is localized and
if the moment is concentrated close to the La. 
 Nuclear magnetic resonance experiments can detect a magnetic moment.
But the magnetic moment is small and for a total moment of 0.19 $\mu_B$ one finds a La contact field
of less than about 40 kG. The spin-splitting is not exceeding 5 mRy within the La valence and core levels,
which is small to be observed directly by spectroscopic methods.
Furthermore, the low La concentration for the real magnetic system might be a problem for the intensity.
The moment is mainly localized to the impurity, with weak interactions with the exterior. This situation
resembles that of a Kondo impurity, but due to the small moment it is questionnable if this can produce the
typical manifestations on specific heat and resistivity at low T. 
Instead of looking for the moment directly, one can propose to use spectroscopic methods
like photoemission, soft X-ray emission spectroscopy (XES) or soft X-ray absorption (SXA), to determine
the degree of localization of the La-band. This can be done at larger La concentrations, 
before magnetism sets in.
Such measurements could reveal
if the DOS shows a localized La-band similar to the DOS calculated here for $x \sim 0.125$ or 
even more at $x \sim 0.037$, or
if the DOS is more like a rigid-band image of the undoped DOS. However, the low concentration of La can be a problem 
in some cases and
lead to a inconclusive result. Soft x-ray photoemission (XPS), for instance, gives a often a good picture of
the total DOS modulated by matrix elements, but if the La concentration is very small one can miss the interesting
part regarding the localized impurity state. XES and SXA should be more useful, since they allow for a local 
projection of the bandstructure by the involvement of a core hole.
 A La core state can be used to probe the occupied or
unoccupied La valence states in XES or SXA, respectively. The signal might be weak because of few La sites, 
but the spectrum
has the advantage of being a selective $\ell$-projection of the local La-DOS.
This means that the signal from La should show a narrow band, while a wide unstructured spectrum (similar to the
total DOS for the undoped material) is expected if a core level from Sr or an In-impurity is used. Dipolar matrix 
elements do some reshaping of the pure DOS functions, but they can be estimated quite easily. 
The question about relaxation around the core hole is more complicated.
According to the so-called final-state rule, one observes the
spectrum of the relaxed, final state \cite{fsr}. 
If so, one should prefer XES, since the relaxed state is the unperturbed ground state,
and the complication of a new calculation including a core-hole for SXA, can be avoided.

Theoretical and experimental efforts have been done to determine the degree of band overlap at the X-point in
the undoped hexaborides \cite{mass,rod,arpes,dhva}. Calculations using gradient corrections to the LDA or using
the GW method, find different values of the band overlap \cite{rod,tro}. 
If the overlap is absent and no doping, the material will be a small-gap semiconductor. 
The excitonic mechanism can be sensitive
to this, and with pressure or measures for changing the $z$-parameter one can think of tuning
the material for optimal properties.  However, the mechanism for magnetism that we have discussed here is not sensitive
to the band structure of the undoped case. It
is very sensitive to the degree of localization
of the La impurity band. If a band calculation gives a too small localization, a wider band with small derivatives
of the DOS, then the moment goes to zero. So, 
unless there is an unknown connection between the band gap for the undoped
band structure and the localization of the La impurity band, we believe that the exact shape of the bands 
in the undoped system is irrelevant for the mechanism for magnetism around a La-impurity.
  But, since we have seen that the results are sensitive to features of the impurity band, it 
also is probable that lattice relaxation around the impurity, including variations
of the $z$-parameter, will have an effect on the magnetic moment.

This work show that the effect of dilute La-doping on the band structure of SrB$_6$ can lead to the appearance
of a magnetic moment around the impurity. Unfortunately, temperature variations cannot be
 studied with the same level of sophistication.
The experimental observation of magnetism is interesting, but also its temperature dependence is unususal as
magnetic ordering can persist up to about 600 K \cite{you}. The Fermi-Dirac smearing is only one part
of the T-depencencies.  It is known that thermal structural (vibrational) disorder is important  
already at 200-300 K. Delicate changes of band structure and the fine details of the DOS due to thermal
disorder can be essential for finding a correct description of the physical properties of a material \cite{fesi}.
This, disorder in the distribution of impurities, disorder of the local moments, non-collinearity 
and orbital magnetism are not considered here, only the electronic excitations given by the Fermi-Dirac 
distribution.
In general, the calculated magnetic moment disappears for $k_BT$ in the range 2-4 mRy
(300-600 K), but not always. One case with the large cell had the largest moments for quite large T, 
which could be related
to a sharp DOS structures a bit beside the optimal position for E$_F$. If this mean-field picture is sufficient,
one would then expect a wide range of ordering temperatures among different samples, like different
samples showing different amplitudes of the moments. The effects of spacial irregularities of the 
impurity distribution of La or other dopants have not been calculated explicitly. But from the separate results
with different relative positions between E$_F$ and the localized La band, one can imagine that
some samples could be apt for large moments at low T, while others with lower moments could resist
better at large T.

Finally, the question is whether this mechanism for magnetism is rare and exotic, or if systems
other than the hexaborides can show similar effects. From eq.  \ref{eq:msto} one should have a small total
DOS, but large DOS derivative of a sub-band. Pure compounds with wide, hybridized bands can hardly
be interesting cases. Dilute doping in a low DOS material seems to give optimal conditions,
because $N_v/N \rightarrow 1$, and an impurity band can localize and become narrow for large distances between
impurities. Not all combinations between impurities and host material will do, because the hybridization
can make the impurity band to merge into the host bands, as for In in the hexaboride. The requirement of
a large $U_0$ with the correct sign, will also reduce the number of candidates for magnetism of this kind.
Preliminary calculations for metal impurities in some oxides show similar results as for the La
doping here, although there is a sizable polarization in the proximity of the impurity \cite{wt1}.

\section{Conclusion.}

We have shown that ferromagnetism occurs in the supercells of LaSr$_7$B$_{48}$ and LaSr$_{26}$B$_{162}$. 
By doing the calculations with different number of k-points and different basis functions,
 it has been possible to show that the
actual derivative of the local DOS at $E_F$, which is necessary for charge transfers, scales with
the stability of the magnetic moment. This demonstrates that the mechanism for ferromagnetism
is not exclusively found in the exchange energy.  The necessary condition of a localized, 
non-constant La-DOS near $E_F$,
appears to be enforced as the doping decreases, and the mechanism of charge transfers persists
at lower doping levels. 
Thus, it is likely that the observed weak ferromagnetism for dilute La-doping in
hexaborides \cite{you,tera}, is a band effect of the kind that we have described. This conclusion is
based on results from first-principle LMTO calculations. The moments are of the order 0.05-0.1 $\mu_B$/La
for the calculations with the "best" choice of basis function and the largest number of k-points. However,
the fact that the results can be "boosted" by choosing somewhat extreme linearization energies, brings up
the problem of describing a very sensitive band structure with a linearized method. Most band results
(total energy, Fermi surface, and the general shape of bands and DOS functions) usually show only a moderate
sensitivity
to the choice of basis or k-point convergence. But here, although the visual inspection of the DOS show
no large differences, there are small differences in the localization of the impurity bands depending
on the choice of basis, which turn out to be important for the size of the magnetic moment.
Therefore, one must emphasize that the calculations show the likelihood of a ferromagnetic state,
but that one can not be affirmative for the absolute values. This situation may reflect the
reality, since different samples are reported to have different moments (at most
0.32 $\mu_B$/La), while some show only large paramagnetic susceptibilities \cite{tera}. Real materials
have some randomness in the distribution of the impurities, while the calculations are made for ordered
supercells. Clustering and other lattice imperfections can be important for the properties 
if they lead to a rigid-band like modification of the band structure, or if the impurity states
become more localized.  From the balance between such effects, it is suggested that
optimal conditions for magnetism can be found indirectly by addition of "rigid-band" dopants like Al.
It is probable that other systems, with localized bands from selected dilute doping in low DOS materials,
can be found to show similar type of weak magnetism.


\begin{acknowledgements}

I am grateful to R. Monnier for discussions, and to H.R. Ott for sending reprints.

\end{acknowledgements}







\begin{table}
\caption{\label{tab:tm}
Paramagnetic DOS values (states/cell Ry, and states/cell Ry$^2$),
  and magnetic moment on La and total ($\mu_B$),
 as function of the number of k-points for the 80-site unit cells. To the impurity site is counted also the contributions
from the 6 closest empty spheres. Values within parentheses are from the calculations with the reduced basis.}
\begin{tabular}{cccccc} 
  k-points & $N(E_F)$ & $N_{La}$ & $N'_{La}$  & $m_{La}$ & $m_{tot}$  \\ 
 10 & 52(54) & 23(24) & 830(830)   & 0.10(0.07)   &  0.15(0.12) \\
 20 & 48(48) & 19(21) & 730(840)   & 0.07(0.06)   &  0.10(0.09) 
\end{tabular}
\end{table}

\begin{table}
\caption{\label{tab:ch} Differences in valence charge (el. per site) 
between doped and undoped SrB$_6$-supercell for selected sites,
calculated from 20 k-points in the small unit cell.  The charge within
the 6 closest empty spheres is assigned to the impurity site. 
There is not much variation among
non-equivalent B sites.} 
\begin{tabular}{ccccc} 
 Basis &  Impurity-site (La or Sr) & Sr near La & Sr far from La & B(average)   \\ 
Large & 1.19 & 0.005 & 0.010 & -0.005  \\
Small & 1.15 & 0.006 & 0.009 & -0.001  
 
\end{tabular}
\end{table}

\begin{table}
\caption{\label{tab:tm2} Paramagnetic DOS  (states/cell Ry),
  and magnetic moment on La and total ($\mu_B$),
 as function of the number of k-points for the 270-site unit cells.
 To the impurity site is counted also the contributions
from the 6 closest empty spheres.   The derivatives of the La-DOS are
 roughly 4 times larger than in the results from the 80-site cell, but the non-linearity is large
 and no precise numbers can be given.  The values within parentheses are for a different linearization,
as is explained in the text. 
 The magnetic moments are calculated at low $k_BT \sim$0.5 mRy, except
 the one for 10 k-points without parenthese, which is for $k_BT$=4 mRy, see the text.}
\begin{tabular}{ccccc} 
  k-points & $N(E_F)$ & $N_{La}$   & $m_{La}$ & $m_{tot}$  \\ 
 4 & 230(220) & 90(80)    & 0.13(0.17)   &  0.19(0.25)   \\
 10 & 165(270) & 55(85)    & 0.04(0.09)   &  0.05(0.17) 
\end{tabular}

\end{table}


\begin{references}

\bibitem{aron} M.C. Aronson, L.J. Sarrao, Z. Fisk, M. Whitton and B.L. Brandt,
Phys. Rev. B{\bf 59}, 4720, (1998).

\bibitem{ott1} H.R. Ott, M. Chernikov, M. Felder, L. Degiorgi, E.G. Moshopoulou, J.L. Sarrao
and Z. Fisk, Z. Phys. B {\bf 102}, 337 (1997);
 
\bibitem{deg} L Degiorgi, E. Felder, H.R. Ott, J.L. Sarrao and Z. Fisk, Phys. Rev. Lett {\bf 79}, 5134, (1997).
 
\bibitem{you} D.P. Young, D. Hall, M.E. Torelli, Z. Fisk, J.L. Sarrao, J.D. Thompson,
H.R. Ott, S.B. Oseroff, R.C. Goodrich and R. Zysler, Nature {\bf 397}, 412 (1999).

\bibitem{tera} T. Terashima, C. Terakura, Y Umeda, N. Kimura, H. Aoki and S. Kunii, J. Phys. Soc. Jap.
{\bf 69}, 2423, (2000).

\bibitem{hase} A. Hasegawa and A Yanase, J. Phys C{\bf 12}, 5431 (1979).

\bibitem{mass} S. Massidda, A. Continenza, T. de Pascale and R. Monnier, Z. Phys. B {\bf 102}, 83, (1997).

\bibitem{rod} C.O. Rodriguez, H. Weht and W.E. Pickett, Phys. Rev. Lett {\bf 84}, 3903, (2000).

\bibitem{tro} H.J. Tromp, P. van Gelderen, P.J. Kelly, G. Brocks and P.A. Bobbert,  Cond-Matt/0011109 (2000).

\bibitem{hexlet} T. Jarlborg, Phys. Rev. Lett. {\bf 85}, 186, (2000).

\bibitem{arpes} J.D. Denlinger, J.A. Clarc, J.W. Allen, G.H. Gweon, D.M. Poirier, C.G. Olson, J.L.Sarrao
and Z. Fisk, Cond-Matt/000922 (2000).

\bibitem{dhva} D. Hall, D. Young, Z. Fisk and R.G. Goodrich, Bull. Am. Phys. Soc. {\bf 44}, 215, (1999).

\bibitem{lda} W. Kohn and L.J. Sham, Phys. Rev. {\bf 140}, A1133, (1965);
O. Gunnarsson and B.I Lundquist, Phys. Rev. B{\bf 13}, 4274, (1976).

\bibitem{hunt} E.C. Stoner, Rep. Prog. Phys. {\bf 11}, 43 (1948);  K.L. Hunt,
Roc. Roy. Soc. A{\bf 216}, 103, (1953).

\bibitem{jan} O. Gunnarsson, J. Phys. F {\bf 6}, 587 (1976);
 J.F. Janak, Phys. Rev. B{\bf 16}, 255, (1977).

\bibitem{c15} T. Jarlborg and A. J. Freeman, Phys. Rev. B {\bf 22}, 2332, (1980).

\bibitem{ric} M.E. Zhitomirsky, T. M. Rice and V.I. Anisimov, Nature {\bf 402}, 251, (1999);
 L. Balents and C.M. Varma, Phys. Rev. Lett {\bf 84}, 1264, (2000);
 V. Barzykin and L.P. Gorkov, Phys. Rev. Lett. {\bf 84}, 2207, (2000).

\bibitem{ort} G. Ortiz, M. Harris and P. Ballone, Phys. Rev. Lett {\bf 82}, 5317, (1999).

\bibitem{tj98} T. Jarlborg, Phys. Rev. B {\bf 58}, 9599, (1998).




\bibitem{lmto} O.K. Andersen, Phys. Rev. B{\bf 12}, 3060 (1975);
T. Jarlborg and G. Arbman, J. Phys. F{\bf 7}, 1635 (1977).

\bibitem{fsr}
U. von Barth and G. Grossmann, Phys. Rev. B{\bf 25}, 5150 (1981).

\bibitem{fesi}
T. Jarlborg, Phys. Rev. B{\bf 59}, 15002 (1999).


\bibitem{wt1}
T. Jarlborg, (unpublished) (2000).

\end{references}
\end{document}